\documentclass[twocolumn,amsmath,nofootinbib,amssymb]{revtex4}

\usepackage{epsfig}
\usepackage{graphicx}
\usepackage{dcolumn}
\usepackage{bm}
\usepackage{color}

\newcommand{\lsim}{\mbox{\raisebox{-.9ex}{~$\stackrel{\mbox{$<$}}{\sim}$~}}}
\newcommand{\gsim}{\mbox{\raisebox{-.9ex}{~$\stackrel{\mbox{$>$}}{\sim}$~}}}

\def\thebiblio#1{
\begin{center}\bf \large References
\end{center}
\list
{[\arabic{enumi}]}{\settowidth\labelwidth{#1.}\leftmargin\labelwidth
 \advance\leftmargin\labelsep
 \usecounter{enumi}}
 \def\newblock{\hskip .11em plus .33em minus -.07em}
 \sloppy
 \sfcode`\.=1000\relax}

\begin{document}

\title{Vector Curvaton without Instabilities}

\author{Konstantinos~Dimopoulos}%
\email{k.dimopoulos1@lancaster.ac.uk}
\affiliation{%
Physics Department, Lancaster University}%
\author{Mindaugas~Kar\v{c}iauskas}%
\email{m.karciauskas@lancaster.ac.uk}
\affiliation{%
Physics Department, Lancaster University}%
\author{Jacques M. Wagstaff}%
\email{wagstafj@exchange.lancs.ac.uk}
\affiliation{%
Physics Department, Lancaster University}%

\date{\today}

\begin{abstract}
A vector curvaton model with a Maxwell kinetic term and varying kinetic 
function and mass during inflation is studied. It is shown that, if light
until the end of inflation, the vector field can generate statistical 
anisotropy in the curvature perturbation spectrum and bispectrum, with the 
latter being predominantly anisotropic. If by the end of inflation the vector 
field becomes heavy, then particle production is isotropic and the vector 
curvaton can alone generate the curvature perturbation. The model does not 
suffer from instabilities such as ghosts and is the only concrete model, to 
date, which can produce the curvature perturbation without direct involvement 
of fundamental scalar fields.
\end{abstract}


\maketitle



There has been a recent outburst of activity in the cosmological implications
of vector fields. 
In particular, the possible contribution of vector fields to the curvature 
perturbation $\zeta$ is receiving growing attention 
\cite{vecurv,nonmin,sugravec}, with emphasis on characteristic signatures such 
as statistical anisotropy in the spectrum and bispectrum of $\zeta$ 
\cite{stanis,fnlanis,stanis+,sugravec2}. 
The forthcoming observations of Planck satellite may well detect statistical 
anisotropy in the near future. On the theoretical side, it is a challenge to 
investigate whether the curvature perturbation can be generated without 
fundamental scalar fields. Even though theories beyond the standard model are
abundant in scalar fields, no scalar field has ever been observed. 
The LHC may or may not confirm the existence of scalar fields. In the latter
case it is imperative to find alternatives for the generation of~$\zeta$.

The pioneering work in Ref.~\cite{vecurv} offered such a possibility by showing
that a massive Abelian vector field alone can, in principle, generate $\zeta$ 
following the curvaton mechanism, originally introduced for scalar fields 
\cite{curv}. The only requirement is that some suitable mechanism breaks 
the conformality of the vector field during inflation, such that an (almost) 
scale invariant spectrum of vector field perturbations is generated. This was
firstly attempted by introducing
a non-minimal coupling to gravity, of the form 
$\frac16 RA^2$ \cite{nonmin}. In another attempt, a non-trivial variation of 
the kinetic function was considered \cite{sugravec}. However, in these early 
works on the vector curvaton, statistical anisotropy was ignored. But, as shown
in Ref.~\cite{stanis}, if particle production of the vector field perturbations
is strongly anisotropic the statistical anisotropy in the spectrum can be
excessive. In this case observational bounds force the vector field 
contribution to $\zeta$ to be subdominant. This was found to be the case for
the $\frac16 RA^2$ model \cite{stanis}. Moreover, the model was criticised
for giving rise to instabilities, such as ghosts \cite{peloso}.
In view of these developments, here we reexamine the model of 
Ref.~\cite{sugravec} taking statistical anisotropy fully into account. 

A massive, Abelian vector field has Lagrangian density 
\begin{eqnarray} &
{\cal L}=-\frac{1}{4}fF_{\rm \mu\nu}F^{\mu\nu}+\frac{1}{2}m^2A_\mu A^\mu\;,
 & 
\label{L}
\end{eqnarray}
where $f$ is the kinetic function
and the field strength tensor is
\mbox{$F_{\mu\nu}=\partial_\mu A_\nu - \partial_\nu A_\mu$}. 
Eq.~(\ref{L}) can be 
the Lagrangian density of a massive gauge field. However, we need not 
restrict ourselves to gauge fields only. If no gauge symmetry is considered the
Maxwell type kinetic term is motivated from being one of the few (three) 
choices \cite{carroll} which avoids introducing instabilities, such as ghosts 
\cite{peloso}.

We assume that the contribution of the vector field to the energy budget of the
Universe during inflation is negligible, i.e. inflation is 
isotropic. We also assume that, during inflation, \mbox{$f=f(t)$} 
and \mbox{$m=m(t)$}.

The equations of motion for the vector field 
perturbations have been 
obtained in Ref.~\cite{sugravec}. In Ref.~\cite{sugravec2} these are expressed 
in terms of the physical (in contrast to comoving) vector field $W_\mu$, whose 
spatial components are 
\mbox{$\mbox{\boldmath $W$}\equiv\sqrt f\,\mbox{\boldmath $A$}/a$},
where $a(t)$ is the scale factor of the FRW metric. To study 
particle production 
we promote the perturbations of the vector field to a quantum operator:
\begin{equation}
\,\!
\hat{\delta\mbox{\boldmath $W$}}\!=
\!\int\!\frac{d^3k}{(2\pi)^
3}\;\sum_\lambda\left[
\mbox{\boldmath $e$}_\lambda
\hat a_\lambda
w_\lambda
e^{i\mbox{\scriptsize\boldmath $k\cdot x$}}\!
+\mbox{\boldmath $e$}^*_\lambda
\hat a^\dag_\lambda
w^*_\lambda
e^{-i\mbox{\scriptsize\boldmath $k\cdot x$}}\right],
\label{expand}
\hspace{-1cm}
\end{equation}
where $k\equiv|\mbox{\boldmath $k$}|$, \mbox{$w_\lambda=w_\lambda(t,k)$} are 
the mode functions and $\lambda=L,R,\parallel$ with $L,R$ denoting
the Left and Right transverse polarisations respectively. The polarisation 
vectors $\mbox{\boldmath $e$}_\lambda(\hat{\mbox{\boldmath $k$}})$ can be 
chosen as
%
\begin{eqnarray} &
\hspace{-.5cm}
e_L\equiv\frac{1}{\sqrt2}(1,i,0),\;\;
e_R=\frac{1}{\sqrt2}(1,-i,0),\;\;
e_\parallel=(0,0,1)\,,
 & 
\label{polar}
\end{eqnarray}
with $\hat{\mbox{\boldmath $k$}}\equiv\mbox{\boldmath $k$}/k$. We consider 
canonical quantisation with \mbox{$\left[\hat a_\lambda(\mbox{\boldmath $k$}),
\hat a^\dag_{\lambda'}(\mbox{\boldmath $k$}')\right]=(2\pi)^3
\delta(\mbox{\boldmath $k$}-\mbox{\boldmath $k$}')\delta_{\lambda\lambda'}$}.
The equations of motion for the mode functions are the same as the ones for the
perturbations of the vector field, because the latter are linear. 
By solving these equations it is found that all the components of the vector
field perturbations obtain a scale invariant spectrum if \cite{sugravec2}
\begin{equation}
f\propto a^{-1\pm 3}\quad{\rm and}\quad m\propto a
\label{fm}
\end{equation}
during inflation and \mbox{$M_*\ll H$}, where \mbox{$M\equiv m/\sqrt f$} is the
mass of the physical vector field and the star denotes the time when the 
cosmological scales exit the horizon. 

In the following, unless otherwise stated, we focus on the case, when
\mbox{$f\propto a^{-4}$}, which has the richest phenomenology. We will also 
assume that \mbox{$f\rightarrow 1$} so that the vector field becomes 
canonically normalised at the end of inflation. Thus, if our vector field is 
indeed a massive gauge field, then the field remains weakly coupled during 
inflation because \mbox{$f\sim 1/e^2$}, where $e$ is the gauge coupling.

The mode functions of the transverse components satisfy the equation
\cite{sugravec2}:
\begin{eqnarray}
 & \left[\partial_t^2
+3H\partial_t
+\left(\frac{k}{a}\right)^2(1+r^2)
\right]
w_+=0\,, &
\label{w+}
\end{eqnarray}
where \mbox{$w_L=w_R\equiv w_+$} because the theory is parity invariant, and
\mbox{$r\equiv aM/k$}, with \mbox{$M\propto a^3$} during 
inflation. The solution to Eq.~(\ref{w+}) is well approximated by 
\cite{sugravec2}
\begin{equation}
\,\!
w_+\!=\!\left\{\begin{array}{lr}
\!\!a^
{-\frac32}
\!\sqrt{\frac{\pi}{4H}}\left[
J_
\frac32
(x)-iJ_
{-\frac32}
(x)\right], & 
\!\!\!x\gsim 1\gg z\,,
\\
\vspace{-.1cm} & \vspace{-.1cm}\\
\!\!\frac{i}{\sqrt{2k}}\left(\frac{H}{k}\right), & 
\hspace{-1cm}
\!\!\!x,z\ll 1,
\\
\vspace{-.1cm} & \vspace{-.1cm}\\
\!\!a^
{-\frac32}\!\sqrt{\frac{\pi}{4H}}\left[
\frac{i}{C}J_
\frac12
(z)+\frac{C}{3}J_
{-\frac12}
(z)
\right], & 
\!\!\!z\gsim 1\gg x,
\end{array}\right.
\hspace{-1cm}
\label{w+solu}
\end{equation}
where \mbox{$C\equiv\sqrt{x^3z}=\,$constant} and 
\begin{equation}
x\equiv k/aH\quad{\rm and}\quad z\equiv M/3H\,,
\label{xz}
\end{equation}
which also implies \mbox{$z\propto a^3$} and
\mbox{$r=3z/x\propto a^4$}. Subhorizon scales correspond to
\mbox{$x>1$}, while the field is light when \mbox{$z\ll 1$} so
scale invariance requires \mbox{$z_*\ll 1$}. 
For Eq.~(\ref{w+solu}) we used the 
vacuum initial condition 
\mbox{\raisebox{-.9ex}{~%
$\stackrel{\mbox{$w_+$}}{\mbox{\tiny $_{x\rightarrow\infty}$}}$~}
$\!\!\!\!=\frac{1}{\sqrt{2k}}e^{ix}$}.

The mode functions of the longitudinal component satisfy the equation
\cite{sugravec2}:
\begin{eqnarray}
\hspace{-.6cm}
& \left[\partial_t^2\!+\!
\left(3+\frac{8}{1+r^2}\right)\!H\partial_t+
\frac{24H^2}{1+r^2}+\left(\frac{k}{a}\right)^2(1+r^2)\right]
w_\|\!=\! 0\,. &
\label{wlong}
\end{eqnarray}

Using the initial condition 
\mbox{$\!\!\!\!$\raisebox{-.9ex}{~%
$\stackrel{\mbox{$w_\|$}}{\mbox{\tiny $_{x\rightarrow\infty}$}}$~}
$\!\!\!\!=\frac{\gamma}{\sqrt{2k}}e^{ix}$},
the solution to the above is well approximated by \cite{sugravec2}
\begin{equation}
\!\,
w_\|\!=\!\left\{\begin{array}{lr}
\!\!
-\frac{ia^
{-\frac72}}{3C^2}\!\left(\frac{k}{H}\right)\!\!\sqrt{\frac{\pi}{4H}}
\left[J_
\frac52(x)-iJ_
{-\frac52}(x)\right]\!, & \!\!\!x\!\gsim\!1\!\gg z,
\\
\vspace{-.1cm} & \vspace{-.1cm}\\
\!\!-\frac{1}{\sqrt{2k}}\left(\frac{H}{k}\right)
\frac{1}{z}\,, & \hspace{-1cm}x,z\!\ll\! 1,
\\
\vspace{-.1cm} & \vspace{-.1cm}\\
\!\!a^
{-\frac32}\!\sqrt{\frac{\pi}{4H}}\left[
\frac{iC}{3}J_
\frac12(z)-\frac{1}{C}J_
{-\frac12}(z)\right]\!, & \!\!\!z\!\gsim\! 1\!\gg\! x,
\end{array}\right.
\hspace{-1cm}
\label{wlongsolu}
\end{equation}
where \mbox{$\gamma\equiv\sqrt{1+\frac{1}{r^2}}$} is the Lorentz boost factor, 
which takes us from the frame with \mbox{{\boldmath $k$}$\,=0$} (where there is
no distinction between longitudinal and transverse components) to that of 
momentum \mbox{{\boldmath $k$}$\,\neq 0$}.

From the above we can obtain the power spectra of the superhorizon 
perturbations as 
\mbox{${\cal P}_\lambda=\frac{k^2}{2\pi^2}|\!\!\!\!$
\raisebox{-.9ex}{~%
$\stackrel{\mbox{$w_\lambda$}}{\mbox{\tiny $_{x\rightarrow 0}$}}$~}
$\!\!\!\!|$}. Eqs.~(\ref{w+solu}) and (\ref{wlongsolu}) suggest that,
when \mbox{$x\ll 1$}, there are two regimes depending whether the vector field
is still light at the end of inflation (\mbox{$z\ll 1$}) or not 
(\mbox{$z\gsim 1$}). If the field is light then the typical values of the 
perturbations are:
\begin{equation}
\delta W_+=\sqrt{{\cal P}_+}=\frac{H}{2\pi}\quad{\rm and}\quad
\delta W_\|=\sqrt{{\cal P}_\|}=\frac{H}{2\pi}\frac{1}{z}\,.
\label{linear}
\end{equation}
Thus, in this case \mbox{${\cal P}_\|\gg {\cal P}_+$} and particle production 
is strongly anisotropic. On the other hand, if the field becomes heavy by the
end of inflation, then Eqs.~(\ref{w+solu}) and (\ref{wlongsolu}) 
can be written as
\begin{eqnarray}
& 2\sqrt{\frac{H}{\pi}}\left(\frac{k}{H}\right)^{3/2}w_+=
\frac{i}{\sqrt z}J_
\frac12(z)+\frac13 x^3\sqrt z J_
{-\frac12}(z)\,, & \\
& 2\sqrt{\frac{H}{\pi}}\left(\frac{k}{H}\right)^{3/2}w_\|=
\frac13 x^3\sqrt z J_
\frac12(z)-\frac{1}{\sqrt z}J_
{-\frac12}(z)\,, & 
\end{eqnarray}
which, in the limit \mbox{$z\gg 1$} and \mbox{$x\rightarrow 0$}, give
\begin{eqnarray}
\hspace{-.7cm}
& w_+\!=\!\frac{i}{\sqrt{2H}}\!\left(\frac{H}{k}\right)^{3/2}\frac{\sin z}{z}
\;\;{\rm and}\;\;
w_\|\!=\!-\frac{1}{\sqrt{2H}}\!\left(\frac{H}{k}\right)^{3/2}\frac{\cos z}{z}.
 &
\label{wbigz}
\end{eqnarray}
Since \mbox{$z\gg 1$}, the frequency of oscillations is very large compared
to the expansion rate $H$ and 
we can use the average
values of the power spectra over many oscillations 
\begin{equation}
\!\,
\overline{{\cal P}_+}=\overline{{\cal P}_\parallel}=
\frac{1}{2z^2}\left(\frac{H}{2\pi}\right)^2\!\Rightarrow\,
\delta W_+=
\delta W_\|
=\frac{H}{2\pi}\frac{1}{z\sqrt 2}\,.
\hspace{-1cm}
\label{oscil}
\end{equation}
Therefore, if the vector field becomes heavy by the end of inflation, 
then particle production is isotropic.

As shown in Ref.~\cite{stanis}, the contribution to the curvature 
perturbation spectrum from a vector curvaton field is
\begin{equation}
{\cal P}_{\zeta_A}(\mbox{\boldmath $k$})=\frac49\frac{\hat\Omega_A^2}{W^2}
\left[{\cal P}_++({\cal P}_\parallel-{\cal P}_+)
(\mbox{\boldmath $\hat W\cdot\hat k$})^2\right],
\label{PzA}
\end{equation}
where 
\mbox{$\mbox{\boldmath $\hat W$}\equiv\mbox{\boldmath $W$}/W$},
\mbox{$W\equiv|\mbox{\boldmath $W$}|$} and
{$\hat\Omega_A\equiv\frac{3\Omega_A}{4-\Omega_A}\sim\Omega_A
\equiv\rho_A/\rho$}, where $\rho_A$ is the density of the vector field. 
As in the case of the scalar curvaton paradigm, Eq.~(\ref{PzA}) should be 
evaluated at the time of decay of the curvaton field. Thus, we see that 
${\cal P}_{\zeta_A}$ is isotropic only when particle production is isotropic
and \mbox{${\cal P}_\|\approx{\cal P}_+$}. In the opposite case, 
${\cal P}_{\zeta_A}$ can be strongly isotropic. If this is so, the vector 
curvaton can only generate a subdominant contribution to $\zeta$, with the 
dominant contribution coming from an isotropic source, such as a scalar field.
The significance of the vector field in this case is 
confined to generating statistical anisotropy in the spectrum and bispectrum.
In our model, the isotropy of 
${\cal P}_{\zeta_A}$ is determined by whether or not the vector field is heavy
by the end of inflation.

Statistical anisotropy in the spectrum can be quantified as \cite{GE}
\begin{equation}
{\cal P}_\zeta(\mathbf{k}) =  {\cal P}_\zeta^{\rm iso}(k) 
\left[1+g\left(
\mbox{\boldmath $\hat d\cdot\hat k$}
\right)^2\right],
\label{Pzg}
\end{equation}
where ${\cal P}_\zeta^{\rm iso}$ is the isotropic part, 
\mbox{\boldmath $\hat d$} is the unit vector along a preferred direction
and we kept only the 
leading term in the anisotropy. Currently, the data provide only a weak upper 
bound on the anisotropy parameter, namely \mbox{$g\lsim 0.3$} \cite{GE}.
Using the $\delta N$ formalism, 
we find~\cite{sugravec2}
\begin{equation}
g =\beta\frac{{\cal P}_{\|} - {\cal P}_{+}}{{\cal P}_{\phi}+\beta 
{\cal P}_{+}}\,,
\label{g}
\end{equation}
where ${\cal P}_{\phi}$ is denotes the spectrum 
due to an isotropic source, such as a scalar field $\phi$, 
which must be dominant if 
\mbox{${\cal P}_\|\not\approx{\cal P}_+$}. We also defined 
\mbox{$\beta\equiv \left(\frac{N_{A}}{N_{\phi}}\right)^2$}, where 
\mbox{$N_\phi\equiv\frac{\partial N}{\partial\phi}$} and
%
\mbox{$N_A\equiv\,|${\boldmath $N_A$}$|$} 
with 
\mbox{$N_A^i\equiv\frac{\partial N}{\partial W_i}$}.
%
$N_\phi$ and $N_A$ quantify how much the elapsed e-folds are affected by 
the perturbations of the scalar and vector fields respectively.

Statistical anisotropy in the bispectrum of $\zeta$ generates anisotropic 
non-Gaussianity, which can be quantified by the non-linearity parameter, as
\begin{equation}
f_{\rm NL}=f_{\rm NL}^{\rm iso}\left(1+g_{\rm NL}
\hat W_\perp^2\right),
\label{gNL}
\end{equation}
in analogy with Eq.~(\ref{Pzg}), where 
\mbox{$\hat W_\perp\equiv|\mbox{\boldmath $\hat W$}_\perp|$}, with 
{\boldmath $\hat W$}$_\perp$ being the projection of {\boldmath $\hat W$}
onto the plane defined by the three \mbox{{\boldmath $k$}$_{1,2,3}$} vectors,
which determine the bispectrum. 

For a parity invariant theory, we have \cite{fnlanis}
\begin{equation}
 \frac{6}{5} f_{\rm NL}=\beta^2 {\cal P}^2_+ \frac{3}{2\hat{\Omega}_A}
\frac{1+(p+\frac{1}{8}\kappa p^2)
\hat W_\perp^2}{({\cal P}_\phi + \beta{\cal P}_+)^2}\,,
 \label{fnlp}
\end{equation}
where \mbox{$\kappa=0$} (\mbox{$\kappa=1$}) in the local (equilateral)
configuration and 
\mbox{$p\equiv({\cal P}_\| - {\cal P}_+)/{\cal P}_+$}.

Let us discuss first the anisotropic case. This is realised if the vector 
field remains light until the end of inflation, i.e. \mbox{$\hat m\ll H_*$},
where $\hat m$ is the final (vacuum) value of the the vector field mass and
$H_*$ denotes the inflationary Hubble scale. If the field is light,
\mbox{${\cal P}_\|\gg{\cal P}_+$}. Thus, the contribution of the vector field 
to $\zeta$ must be subdominant. This means that \mbox{$\beta\ll 1$} since
the number of inflationary e-folds is primarily modulated by the scalar field.
Hence, the anisotropy in the spectrum is
\mbox{$g\simeq\beta{\cal P}_\|/{\cal P}_+=\beta/z^2$}. For the 
bispectrum we have \mbox{$f_{\rm NL}^{\rm iso}=2\beta^2/\Omega_A$}, where
we considered \mbox{$\Omega_A\ll 1$} and took 
\mbox{${\cal P}_\phi=\frac{H_*}{2\pi}={\cal P}_+$}. Since 
\mbox{$p=z^{-2}\gg 1$}
the anisotropic part of the bispectrum peaks in the equilateral configuration,
where we find \mbox{$g_{\rm NL}\simeq\frac18 p^2\gg 1$}. Therefore, 
non-Gaussianity is predominantly anisotropic in this case.

Let us now discuss the (almost) isotropic case. This corresponds to 
\mbox{$\hat m\gsim H_*$}, i.e.~the vector field becomes heavy and begins 
oscillating by the end of inflation. In this case, 
\mbox{${\cal P}_\|\approx{\cal P}_+$} and excessive anisotropy in 
${\cal P}_{\zeta_A}$ is avoided. Thus, we can dispense with the 
scalar field and consider \mbox{${\cal P}_\zeta={\cal P}_{\zeta_A}$},
i.e.~the vector curvaton {\em alone} generates $\zeta$. Consequently,
\mbox{$N_\phi\rightarrow 0$}, i.e.~\mbox{$\beta\gg 1$}. Then, from 
Eqs.~(\ref{g}) and (\ref{fnlp}),
\mbox{$g\simeq p\simeq g_{\rm NL}$}, i.e.~statistical anisotropy is equal
in the spectrum and bispectrum and can be non-negligible only if
\mbox{$\hat m\sim H_*$}. If~\mbox{$\hat m\gg H_*$} then 
\mbox{$g=g_{\rm NL}=0$} and $\zeta$ is isotropic. In this 
case, \mbox{$f_{\rm NL}=f_{\rm NL}^{\rm iso}=5/4\hat\Omega_A$} 
as in the scalar curvaton scenario~\cite{curv}.

From Eq.~(\ref{Pzg}) it is evident that the magnitude of the contribution to 
curvature perturbation from the vector curvaton depends on the value of the 
zero mode $W$. It is easy to show that the equation of motion of $W$, under the
conditions in Eq.~(\ref{fm}), is \cite{sugravec2}
\begin{equation}
\ddot W+3H\dot W+M^2W=0
\label{WEoM}
\end{equation}
during inflation, where the dot denotes time derivative. 
When \mbox{$f\propto a^{-4}$} (i.e. \mbox{$M\propto a^3$}) 
and assuming initial energy equipartition between the kinetic and the mass 
terms in Eq.~(\ref{L}), we have \cite{sugravec2}
\begin{equation}
W=W_{0}
\left(\frac{a}{a_{0}}\right)^{-3}
\sqrt{2}\,\cos\left(z\pm\frac{\pi}{4}\right),
\label{W}
\end{equation}
where `0' denotes initial values. 
Thus, 
the typical value of the zero mode scales as \mbox{$W\propto a^{-3}$}.
From the above it can also be shown that the density of the vector field 
remains constant during inflation \mbox{$\rho_A\sim M_0^2W_0^2$} 
\cite{sugravec2}. 

After inflation, $f$ and $m$ assume their final (vacuum) values and 
the equation of motion for $W$ 
becomes \cite{sugravec2}.
\begin{equation}
\ddot W+3H\dot W+(\dot H+2H^2+\hat m^2)W=0\,.
\label{Weom}
\end{equation}
From this, we can deduce that, if the field is still light after inflation,
then \mbox{$\rho_A\propto a^{-4}$}, i.e. $\rho_A$ is diluted as radiation, in 
contrast to the case of a light scalar field whose density remains constant.
When \mbox{$\hat m> H(t)$}, the field becomes heavy and oscillates with 
frequency $\hat m$. Then, one can show that, on average, 
\mbox{$\rho_A\propto a^{-3}$} and \mbox{$P_i=0$} \cite{sugravec2}, i.e. the 
field acts as pressureless {\em isotropic} matter and can dominate the 
Universe without generating excessive large-scale anisotropy, in accordance to
the findings of Ref.~\cite{vecurv}. Thus, it can safely play the role of the 
curvaton.

Working as in Ref.~\cite{nonmin} and assuming prompt reheating, we obtain
\cite{sugravec2}
\begin{equation}
\frac{H_*}{m_P}\sim\Omega_{\rm dec}^{1/2}\;\zeta_{A}\,
\left(\frac{\max\left\{\Gamma_A;H_{\rm dom}\right\}}{
\min\left\{\hat{m};H_*\right\}}\right)^{1/4},
\label{Hcurv}
\end{equation}
where \mbox{$\Omega_{\rm dec}$} is $\Omega_A$ at the time of the vector field 
decay, $\zeta_A$ is the curvature perturbation attributed to the vector field,
$H_{\rm dom}$ is the Hubble scale when the oscillating vector field dominates 
the Universe, if this occurs before decay, and \mbox{$\Gamma_A=h^2\hat m$} is
its decay rate, with $h$ being the coupling to the decay products. The latter
ranges as \mbox{$\hat m/m_P\lsim h\lsim 1$}, where the lower bound 
corresponds to decay through gravitational couplings.

Let us consider first the (almost) isotropic case, where 
\mbox{$\hat m\gsim H_*$}. Then, \mbox{$\zeta\sim\Omega_{\rm dec}\zeta_A$}
as in the scalar curvaton scenario. This implies 
\mbox{$\Omega_{\rm dec}H_*\gsim\zeta^2m_P$}.
Thus, we obtain
\begin{equation}
H_*\gsim 10^9\;{\rm GeV}\;,
\label{Hbound1}
\end{equation}
where we used \mbox{$\zeta=5\times 10^{-5}$}.
Now, because \mbox{$M\propto a^3$} during inflation, we have 
\mbox{$\hat m=e^{3N_{\rm osc}}H_*$}, where $N_{\rm osc}$ is the remaining
e-folds of inflation at the onset of the vector field oscillations. Using this
and considering gravitational decay, it can be shown that
\mbox{$N_{\rm osc}^{\rm max}=-\frac49\ln\zeta=4.4$} \cite{sugravec2}. 
Thus, the parameter space for this scenario is
\begin{equation}
1\lsim \hat m/H_*< 10^6.
\label{mbound1}
\end{equation}
The above range is somewhat reduced if the decay 
of the curvaton is more efficient than gravitational.
Indeed, when \mbox{$h\sim 1$}
the upper bound is reduced by 
${\cal O}(10^2)$ since
{$N_{\rm osc}^{\rm max}=-\frac23\ln\Omega_{\rm dec}\lsim 3.1$},
where \mbox{$\Omega_{\rm dec}\gsim 10^{-2}$} because a smaller
$\Omega_{\rm dec}$ violates the current observational bounds on
\mbox{$f_{\rm NL}=5/4\hat\Omega_{\rm dec}$}. Therefore, the 
parameter space where the vector field undergoes isotropic particle 
production and can alone account for 
$\zeta$ is exponentially large.

Let us consider now the anisotropic case, where \mbox{$\hat m\ll H_*$}. 
Then,
it can be shown that
\mbox{$\zeta\sim g^{-1/2}\Omega_{\rm dec}\zeta_A$} \cite{sugravec2}. 
A lower bound on $\hat m$ is obtained due to the requirement
that the vector curvaton decays before big bang nucleosynthesis 
\mbox{$\Gamma_A>T_{\rm BBN}^2/m_P$}, where \mbox{$T_{\rm BBN}\sim 1\;$MeV}. 
Hence, we get
\begin{equation}
H_{*}>\sqrt g\;10^7\;{\rm GeV}\;,
\label{Hbound2}
\end{equation}
while the parameter space for this scenario is
\begin{equation}
10\;{\rm TeV}\lsim\hat m\ll H_*\;.
\label{mbound2}
\end{equation}

Before concluding let us briefly discuss also the case when 
\mbox{$f\propto a^2$} in which too a scale invariant spectrum of all the
components of the vector field perturbations is attained. In this case 
\mbox{$M=\;$constant} throughout inflation. Thus, 
we have 
\mbox{$\hat m=M_*\ll H_*$}. It turns out that Eq.~(\ref{linear}) is also valid 
in this case, but crucially \mbox{$1\gg z=\;$constant}, which means that 
\mbox{$1\ll {\cal P}_\|/{\cal P}_+=\;$constant}.
Thus, particle production is always 
strongly anisotropic. The predictions for $g$ and $g_{\rm NL}$ are identical 
with the case when when \mbox{$f\propto a^{-4}$}. The zero mode satisfies
Eq.~(\ref{WEoM}) but initial energy equipartition suggests that
\mbox{$W=W_0=\;$constant} during inflation, which also implies that
\mbox{$\rho_A\sim M_0^2W_0^2$}. 
Since the evolution after inflation is insensitive to the scaling of $f$ during
inflation, the bounds from 
the curvaton mechanism are the 
same with those in the anisotropic case discussed above for 
\mbox{$f\propto a^{-4}$} (e.g. Eqs.~(\ref{Hbound2}) and (\ref{mbound2})
remain valid). All in all 
the predictions of our model for \mbox{$f\propto a^{-1\pm 3}$} are identical, 
when \mbox{$\hat m\ll H_*$}, while the possibility that 
\mbox{$\hat m\gsim H_*$} can only be realised when \mbox{$f\propto a^{-4}$}.

In conclusion, we have presented and analysed a new model of vector curvaton 
field, which does not suffer from instabilities such as ghosts. The model
corresponds to a massive Abelian vector field with a Maxwell type kinetic term,
whose kinetic function and mass scale during inflation according to 
Eq.~(\ref{fm}). The implications of our model depend on whether the field is 
light or not by the end of inflation 
If it is light, then particle production is anisotropic. In this case, the 
vector curvaton's contribution to the curvature perturbation must be 
subdominant with the dominant part due to some isotropic source such as a 
scalar field. However, the vector field can generate statistical anisotropy in 
the spectrum, which could be observable. It also generates anisotropic 
non-Gaussianity along a direction correlated with the statistical anisotropy in
the spectrum. In fact, the anisotropy in the bispectrum is dominant to the
isotropic part, which means that, if a non-Gaussian signal without angular 
modulation is indeed observed e.g. by the Planck mission, our model will be 
falsified in this case. However, in the case when the field becomes heavy by 
the end of inflation, particle production becomes isotropic. Then, the vector 
curvaton's contribution to the curvature perturbation can be dominant. Indeed, 
we can dispense with contributions from other sources such as scalar fields and
consider that {\em the vector curvaton alone generates 
the curvature perturbation in the Universe}.
If there is some residual statistical anisotropy, we have shown that its 
magnitude is the same in both the spectrum and bispectrum, which is a 
characteristic signature of this possibility. However, if the vector field is
much heavier than the Hubble scale by the end of inflation, any anisotropy is
negligible and the implications of the vector curvaton are indistinguishable
to the scalar curvaton scenario. We have found that there is ample parameter 
space for both the above cases (see Eqs. (\ref{mbound1}) and (\ref{mbound2})).

The curvature perturbation spectrum obtained is exactly flat. We can get
a red spectrum (currently favoured by the data) by assuming that inflation is 
not exactly de Sitter, with 
\mbox{$\varepsilon\equiv -\dot H/H^2\sim 10^{-2}$}. The model needs to be 
tuned according to Eq.~(\ref{fm}), but this has to be contrasted with the
unavoidable fine-tunning of scalar field models ($\eta$-problem). 
Our model may be realised in the context of supergravity (supermassive gauge 
fields) or superstrings (vector fluxes in the bulk).

This work was supported (in part) by the 
UniverseNet Network MRTN-CT-2006-035863 
and by STFC Grant ST/G000549/1.
M.K. and J.M.W. are also supported by the Lancaster University Physics 
Department.

\begin{thebiblio
}{99}

\bibitem{vecurv}
K.~Dimopoulos,
  Phys.\ Rev.\  D {\bf 74} (2006) 083502.

\bibitem{nonmin}
K.~Dimopoulos and M.~Kar\v{c}iauskas,
  JHEP {\bf 0807} (2008) 119.

\bibitem{sugravec}
K.~Dimopoulos,
  Phys.\ Rev.\  D {\bf 76} (2007) 063506.

\bibitem{stanis}
K.~Dimopoulos, M.~Kar\v{c}iauskas, D.~H.~Lyth and Y.~Rodriguez,
  JCAP {\bf 0905} (2009) 013.

\bibitem{fnlanis}
 M.~Kar\v{c}iauskas, K.~Dimopoulos and D.~H.~Lyth,
  Phys.\ Rev.\  D {\bf 80} (2009) 023509.

\bibitem{stanis+}
S.~Yokoyama and J.~Soda,
  JCAP {\bf 0808} (2008) 005;
S.~Kanno, M.~Kimura, J.~Soda and S.~Yokoyama,
  JCAP {\bf 0808} (2008) 034;
M.~a.~Watanabe, S.~Kanno and J.~Soda,
  Phys.\ Rev.\ Lett.\  {\bf 102} (2009) 191302;
N.~Bartolo, E.~Dimastrogiovanni, S.~Matarrese and A.~Riotto,
0906.4944 [astro-ph.CO].

\bibitem{sugravec2}
K.~Dimopoulos, M.~Karciauskas and J.~M.~Wagstaff,
0907.1838 [hep-ph].


\bibitem{curv}
D.~H.~Lyth and D.~Wands,
Phys.\ Lett.\ B {\bf 524} (2002) 5;
T.~Moroi and T.~Takahashi,
Phys.\ Lett.\ B {\bf 522}, 215 (2001)
[Erratum-ibid.\ B {\bf 539}, 303 (2002)];
K.~Enqvist and M.~S.~Sloth,
Nucl.\ Phys.\ B {\bf 626} (2002) 395.

\bibitem{peloso}
B.~Himmetoglu, C.~R.~Contaldi and M.~Peloso,
  Phys.\ Rev.\  D {\bf 79} (2009) 063517;
  Phys.\ Rev.\ Lett.\  {\bf 102} (2009) 111301.

\bibitem{carroll}
S.~M.~Carroll, T.~R.~Dulaney, M.~I.~Gresham and H.~Tam,
  Phys.\ Rev.\  D {\bf 79} (2009) 065011;
T.~R.~Dulaney, M.~I.~Gresham and M.~B.~Wise,
  Phys.\ Rev.\  D {\bf 77} (2008) 083510
  [Erratum-ibid.\  D {\bf 79} (2009) 029903].

\bibitem{GE}
N.~E.~Groeneboom and H.~K.~Eriksen,
  Astrophys.\ J.\  {\bf 690} (2009) 1807.

\end{thebiblio}


\end{document}